\documentclass[%%
a4paper,%           % letterpaper, a4paper, a5paper               %
10pt,%               % 10pt,11pt, 12pt                             %
superscriptaddress,% % authors with affiliations via superscripts  %
amsmath,%            % add AMS-Latex features                   	 %
amssymb,%            % add extra AMS symbols, including amsfonts   %
aps,%				 %												 %
prd,%       	     % prl, pra, prb, prc, prd, pre, prstab        %
showpacs,% 			 % make PACS codes appear                      %
eqsecnum,%          % number equations by section              	 %
]{revtex4}%         %                                           	 %

\usepackage{tensor}      % manipulate tensors				       %
\usepackage{graphicx}    % include figures  					   %
\usepackage[colorlinks=true,%        color link				   %
citecolor=blue,%		  % cite color							   %
linkcolor=blue,%		  % link color							   %
urlcolor=blue%			  % url color							   %
]{hyperref}				  % create hyperlinks		 			   %
\usepackage{bm}		      % \bm{<text>} Bold math symbols   	   %
\usepackage{dcolumn}	  % Align table columns on decimal point  %
\usepackage{array}	  	  % dcolumn depends on array			   %
\newcommand*{\be}{\begin{equation}}
\newcommand*{\ee}{\end{equation}}
\newcommand*{\bea}{\begin{eqnarray}}
\newcommand*{\eea}{\end{eqnarray}}

\begin{document}

%%%%%%%%%%%%%%%%%%%%%%%%%%%%%%%%%%%%%%%%%%%%%%%%%%%%%%%%%%%%%%%%%%%%%%
%%%%%%%%%%%%%%%%%%%%%%%%%%% the front matter %%%%%%%%%%%%%%%%%%%%%%%%%
%%%%%%%%%%%%%%%%%%%%%%%%%%%%%%%%%%%%%%%%%%%%%%%%%%%%%%%%%%%%%%%%%%%%%%
\begin{titlepage}

\begin{flushright}
 arXiv:1410.7715
\end{flushright}

\title{\Large \bf Post-Newtonian Approximation of
Teleparallel~Gravity~Coupled~with a Scalar Field}

%%%%%%%%%%%%%%%%%%%%%%%%%%%%%%%%%%%%%%%%%%%%%%%%%%%%%%%%%%%%%%%%%%%%%%
%%%%%%%%%%%%%%%%%%%%%%%%%%%%%%% authors %%%%%%%%%%%%%%%%%%%%%%%%%%%%%%
\author{Zu-Cheng~Chen\,}
\email[\,Email address:\ ]{bingining@gmail.com}
\affiliation{School of Physics,
Beijing Institute of Technology, Beijing 100081, China}

\author{You~Wu\,}
\email[\,Email address:\ ]{5u@ruc.edu.cn}
\affiliation{Department of Physics,
Renmin University of China,
Beijing 100872, China}

\author{Hao~Wei\,}
\thanks{\,Corresponding author}
\email[\,Email address:\ ]{haowei@bit.edu.cn}
\affiliation{School of Physics,
Beijing Institute of Technology, Beijing 100081, China}

%%%%%%%%%%%%%%%%%%%%%%%%%%%%%%%%%%%%%%%%%%%%%%%%%%%%%%%%%%%%%%%%%%%%%%
%%%%%%%%%%%%%%%%%%%%%%%%%%%%% abstract %%%%%%%%%%%%%%%%%%%%%%%%%%%%%%%
\begin{abstract}\vspace{1cm}
\centerline{\bf ABSTRACT}\vspace{2mm}
We use the parameterized post-Newtonian (PPN) formalism to
explore the weak field approximation of teleparallel gravity
non-minimally coupling to a scalar field $\phi$, with arbitrary
coupling function $\omega(\phi)$ and potential $V(\phi)$. We
find that all the PPN parameters are identical to general
relativity (GR), which makes this class of theories compatible
with the Solar System experiments. This feature also makes the
theories quite different from the scalar-tensor theories, which
might be subject to stringent constraints on the parameter
space, or need some screening mechanisms to pass the Solar
System experimental constraints.
\end{abstract}

% see http://www.aip.org/pacs/

\maketitle

\end{titlepage}

%%%%%%%%%%%%%%%%%%%%%%%%%%%%%%%%%%%%%%%%%%%%%%%%%%%%%%%%%%%%%%%%%%%%%%
%%%%%%%%%%%%%%%%%%%%%%%%%%%%% section 1 %%%%%%%%%%%%%%%%%%%%%%%%%%%%%%
%%%%%%%%%%%%%%%%%%%%%%%%%%%%%%%%%%%%%%%%%%%%%%%%%%%%%%%%%%%%%%%%%%%%%%
\section{\label{Intro}Introduction}
Since the late-time acceleration of our universe
has been confirmed by various observations
\cite{Riess:1998cb,Perlmutter:1998np,Spergel:2003cb,%%
Tegmark:2003ud,Allen:2004cd,Riess:2004nr,Spergel:2006hy,%%
Ade:2013ktc,Ade:2013zuv}, in the literature
there are much efforts to explain this surprising
phenomenon. Although it is straightforward to introduce a
cosmological constant \cite{Weinberg:1988cp,%%
Carroll:1991mt,Zlatev:1998tr,Carroll:2000fy,Peebles:2002gy,%%
Padmanabhan:2002ji,Sola:2013gha} to account for the acceleration, it
also gives rise to the so-called fine-tuning \cite{Weinberg:1988cp}
and cosmic coincidence \cite{fitch1997critical} problems.
Other kinds of dark energy (DE)
\cite{Copeland:2006wr,Weinberg:2012es,wang2010dark},
e.g. quintessence \cite{Caldwell:1997ii,Ford:1987de,%%
Ratra:1987rm}, phantom \cite{Caldwell:1999ew,Caldwell:2003vq},
$k$-essence \cite{ArmendarizPicon:2000ah,Chiba:1999ka,%%
ArmendarizPicon:1999rj,Garriga:1999vw},
and tachyon \cite{Copeland:2004hq,Bagla:2002yn,
Gibbons:2002md,Abramo:2003cp,Padmanabhan:2002cp},
work as well, but one also needs to figure out
why it is homogeneous \cite{Vachaspati:1998dy}
and why it has recently achieved dominance
\cite{Carroll:2000fy,Peebles:2002gy,Padmanabhan:2002ji,%%
Copeland:2006wr}. On the other hand, because the energy scale
of the field potential is very low, it is difficult to
construct viable scalar field models in the framework of
particle physics. Instead of assuming the existence of a
mysterious DE with exotic properties, an alternative approach
is to modify Einstein's general relativity (GR) on the
cosmological scales while GR can be restored on small scales.
In the literature, such approaches are usually called modified
gravity theories \cite{Clifton:2011jh,tsujikawa2010modifieda}.
In particular, the scalar-tensor theories
\cite{Bergmann:1968ve,Wagoner:1970vr,EspositoFarese:2000ij,%%
cosmology,fujii2004scalartensor}
which introduce an extra degree of freedom, namely a scalar
field $\phi$, coupling to the gravitational sector (the Ricci
scalar $R$), might be one of the most natural alternatives to
GR, since such a scalar field generically could arise in
the attempts to quantize gravity (e.g. string theory).
Scalar-tensor theories can not only describe the deviation
from GR to have the desired cosmological dynamics on large
scales \cite{Bartolo:1999sq,Boisseau:2010pd,Charmousis:2011bf},
but they also respect most of the GR's
symmetries, e.g. the local Lorentz invariance.

On the other hand, torsion tensor can naturally arise
when one studies the gauge theories which try to quantize
gravitational field and unify it with other fundamental
interactions. In fact, spin and torsion can be formulated
naturally and elegantly in such gauge formulations of gravity
\cite{hehl1976generala,hehl1980gravitation}.
By introducing the curvatureless Weitzenb\"ock connection
\cite{weitzenbock1923invariantentheorie} instead of the
torsionless Levi-Civita connection used in GR, the so-called
Teleparallel Equivalent of General Relativity (TEGR), or also
known as teleparallel gravity, can be formulated,
which naturally arises within the framework of the gauge theory
of the spacetime translation group. Teleparallel gravity uses
the vierbein field as the basic dynamical quantity instead of
metric in GR, and attributes gravitation to the torsion tensor.
After it was originally proposed by Einstein in 1920s
\cite{einstein1928riemann,einstein1928neue,%%
einstein1930auf,Unzicker:2005in}, teleparallel gravity has
been extensively studied in the literature
(see e.g. \cite{hayashi1967extendeda,cho1976einstein,%%
cho1976gauge,hayashi1977gauge,hayashi1979new}).
As is well known, the Friedmann-Robertson-Walker (FRW) universe
in the framework of teleparallel gravity is completely
equivalent to a matter-dominated universe in the framework of
GR, and hence cannot be accelerated. In the literature, there
are two ways out. In analogy to the well-known $f(R)$ theory,
the first approach is to generalize teleparallel gravity to the
so-called $f(T)$ theory \cite{Bengochea:2008gz,Linder:2010py}.
The second approach is to directly add DE into teleparallel
gravity. Of course, the simplest candidate of DE
is quintessence characterized by a canonical
scalar field. Inspired by the well-known scalar-tensor
theories, it is reasonable to introduce a non-minimal coupling
between the scalar field and the torsion scalar $T$. The
so-called teleparallel dark energy \cite{Geng:2011aj,%%
Geng:2011ka,Wei:2011yr,Xu:2012jf,Geng:2012vn,Geng:2013uga,%%
Gu:2012ww,Li:2013oef}, in which the canonical scalar field
(quintessence) is coupled with the gravitation, has been shown
that it can drive the cosmic acceleration even when the potential of
the scalar field vanishes \cite{Geng:2013uga,Gu:2012ww}. Note that
in e.g. \cite{Geng:2011aj,Geng:2011ka,Wei:2011yr,Xu:2012jf,%%
Geng:2012vn,Geng:2013uga,Gu:2012ww,Li:2013oef} the coupling is
chosen to be a particular form. Later, the teleparallel dark energy
model has been generalized in various directions. For instance, the
so-called tachyonic teleparallel dark energy model, in which a
non-canonical scalar field (tachyon field) is coupled with
gravitation, has been shown that the effective
equation-of-state parameter (EoS) of DE can cross the phantom
divide, and the cosmological coincidence problem could be alleviated
\cite{Banijamali:2012nx,Otalora:2013dsa,Fazlpour:2014qaa,%%
Banijamali:2014nba}. Noether symmetry has been studied
\cite{Kucukakca:2013mya,Kucukakca:2014vja} in the teleparallel dark
energy model, in which the coupling constant is extended to be
a general coupling function. It is claimed that the effective
EoS can cross the phantom divide if the coupling function and
the potential of the scalar field are of power-law forms.

No matter how successful an alternative theory to GR is on the
cosmological scales, it should also have the appropriate
Newtonian and post-Newtonian approximations in order to pass
the local tests in Solar System. As is well known, a natural
framework to test the weak field limit of a gravity theory is
given by the parameterized post-Newtonian (PPN) formalism (see
e.g.~\cite{will1993theorya}). In fact, modified gravity
theories are usually subject to much severer constraints from
the Solar System experiments than the ones from cosmological
observations. For instance, the parameter of the prototypical
Brans-Dicke theory $\omega_{\rm BD}$ \cite{brans1961machs} was
constrained to $\omega_{\rm BD} > 181.65$ at $2\sigma$
confidence level by using Planck data of the cosmic microwave
background (CMB) combined with the baryon acoustic oscillation
(BAO) data in \cite{Li:2013nwa}, and $\omega_{\rm BD} > 40000$
at $2\sigma$ confidence level by using the tracking data
obtained from the Cassini mission \cite{Bertotti:2003rm}. On
the other hand, some types of modified gravity theories are
even claimed to be incompatible with the local tests in Solar
System \cite{Soussa:2003re,Chiba:2003ir,Olmo:2005zr}, and
hence cannot be viable candidates to explain the cosmic
acceleration. In the more general scalar-tensor theories and
$f(R)$ theories, the well-known Chameleon mechanism is invoked
to screen the fifth force \cite{Khoury:2003rn,Khoury:2003aq,%%
Tsujikawa:2008uc,Kase:2013uja}, and hence they have no
significant deviation from GR on small scale, while they can
still drive the acceleration of the universe on cosmological scale.
Similarly, the Vainshtein mechanism
\cite{Vainshtein:1972sx,Deffayet:2001uk,ArkaniHamed:2002sp} and
the Symmetron mechanism
\cite{Hinterbichler:2010es,Hinterbichler:2011ca} are also
extensively invoked in other types of modified gravity theories
to pass the local tests in Solar System.

Motivated by the above discussions, it is necessary and worth
to explore the weak field behaviors of modified gravities.
Recently, the PPN parameters for the teleparallel dark energy
model have been explicitly calculated in \cite{Li:2013oef}, and
it is claimed that the potential of the scalar field has no
effect on PPN parameters and hence this model can be compatible
with the local tests in Solar System. Note that in \cite{Li:2013oef}
the coupling is chosen to be a particular form. In the present
work, we try to generalize the work of \cite{Li:2013oef} and
explore the weak field approximation of teleparallel gravity
non-minimally coupling to a scalar field $\phi$ with arbitrary
coupling function $\omega(\phi)$ and potential $V(\phi)$, by
explicitly calculating the corresponding PPN parameters. This
paper is organized as follows. We give a brief review of
teleparallel gravity in Sec.~\ref{TeleGra}. Next, we present
the action functional for the teleparallel gravity coupled
with a scalar field and derive the corresponding field
equations in Sec.~\ref{TeleGraS}. We then expand the field
equations to sufficient orders and solve the perturbations to
obtain the post-Newtonian approximation in Sec.~\ref{PNA}.
Finally, some concluding remarks are given in Sec.~\ref{Conclusions}.

%%%%%%%%%%%%%%%%%%%%%%%%%%%%%%%%%%%%%%%%%%%%%%%%%%%%%%%%%%%%%%%%%%%%%%
%%%%%%%%%%%%%%%%%%%%%%%%%%%%% section 2 %%%%%%%%%%%%%%%%%%%%%%%%%%%%%%
%%%%%%%%%%%%%%%%%%%%%%%%%%%%%%%%%%%%%%%%%%%%%%%%%%%%%%%%%%%%%%%%%%%%%%
\section{\label{TeleGra}Teleparallel Gravity}

Here we give a brief review of teleparallel gravity.
Teleparallel gravity uses a vierbein field
$e_a = \tensor{e}{_a^\mu} \partial_{\mu}$ as dynamical
quantity, with Latin indices $a, b, \cdots = 0, \cdots, 3$,
and $i, j, \cdots = 1, \cdots, 3$, Greek indices
$\mu, \nu, \cdots = 0, \cdots ,3$, and $\partial_{\mu}$
coordinate bases. We also note that the Einstein summation
notation for the indices is used throughout this work.
The vierbein is an orthonormal basis for
the tangent space at each point $x^{\mu}$ of the manifold,
namely $e_a \cdot e_b = \eta_{a b}$, with
$\eta_{a b} = \mathrm{diag}\,(-1,\, 1,\, 1,\, 1)$.
Then the metric tensor can  be expressed in the dual vierbein
$\tensor{e}{^a_\mu}$ as
\be\label{gTe}
	g_{\mu \nu}(x)
	= \eta_{a b}\, \tensor{e}{^a_\mu}(x)\, \tensor{e}{^b_\nu}(x) .
\ee
Rather than using the torsionless Levi-Civita connection in GR,
teleparallel gravity uses the Weitzenb\"{o}ck connection
$\tensor{\Gamma}{^{\lambda}_{\mu}_{\nu}}$
\cite{weitzenbock1923invariantentheorie}, which is defined by
\be
	\tensor{\Gamma}{^\lambda_\mu_\nu}
	= \tensor{e}{_a^\lambda}\, \partial_{\mu} \tensor{e}{^a_\nu} .
\ee
Note that the lower indices $\mu$ and $\nu$ are not symmetric
in general, thus the torsion tensor (will be defined below)
is non-vanishing in the teleparallel spacetime.
The Weitzenb\"{o}ck torsion tensor is defined by
\be
	\tensor{T}{^\lambda_\mu_\nu}
	= \tensor{\Gamma}{^\lambda_\nu_\mu}
		- \tensor{\Gamma}{^\lambda_\mu_\nu}
	= \tensor{e}{_a^\lambda} \left(
			\partial_\nu \tensor{e}{^a_\mu}
	 		- \partial_\mu \tensor{e}{^a_\nu}
	 		\right) .
\ee
In teleparallel gravity, the gravitational action is given by
the torsion scalar instead of the Ricci scalar in GR.
The torsion scalar is basically the square of the
Weitzenb\"{o}ck torsion tensor, and reads
\be
	T = \tensor{S}{^\rho_\mu_\nu} \tensor{T}{_\rho^\mu^\nu}
	  = \frac{1}{4} \tensor{T}{^\rho_\mu_\nu} \tensor{T}{_\rho^\mu^\nu}
		+ \frac{1}{2} \tensor{T}{^\rho_\mu_\nu}
			\tensor{T}{^\nu^\mu_\rho}
		- \tensor{T}{^\rho_\mu_\rho} \tensor{T}{^\nu^\mu_\nu} ,
\ee
with the super-potential tensor $\tensor{S}{^\rho_\mu_\nu}$
defined by
\be
	\tensor{S}{^\rho_\mu_\nu}
	  = \frac{1}{4} \left(
	  		\tensor{T}{^\rho_\mu_\nu}
			- \tensor{T}{_\mu_\nu^\rho}
			+ \tensor{T}{_\nu_\mu^\rho}
			\right)
		+ \frac{1}{2} \delta^\rho_\mu \tensor{T}{^\sigma_\nu_\sigma}
		- \frac{1}{2} \delta^\rho_\nu \tensor{T}{^\sigma_\mu_\sigma} .
\ee
The gravitational field is driven by the torsion scalar $T$,
and the action reads
\be
	S = \frac{1}{2 \kappa^2} \int e T d^4x
		+ S_m\left[ \tensor{e}{_a^\mu}, \chi_m \right] ,
\ee
where $e = \mathrm{det}\,(\tensor{e}{^a_\mu}) = \sqrt{-g}$ and
$\kappa^2 = 8 \pi G_N$, with $g$ the determinant of the metric
$g_{\mu\nu}$ and $G_N$ the Newtonian constant. Note that we
have used the units in which the speed of light $c = 1$, and
the reduced Planck constant $\hbar = 1$.
$S_m\left[ \tensor{e}{_a^\mu}, \chi_m \right]$ is the matter part
of the action, and $\chi_m$ denotes all matter fields collectively.

%%%%%%%%%%%%%%%%%%%%%%%%%%%%%%%%%%%%%%%%%%%%%%%%%%%%%%%%%%%%%%%%%%%%%%
%%%%%%%%%%%%%%%%%%%%%%%%%%%%% section 3 %%%%%%%%%%%%%%%%%%%%%%%%%%%%%%
%%%%%%%%%%%%%%%%%%%%%%%%%%%%%%%%%%%%%%%%%%%%%%%%%%%%%%%%%%%%%%%%%%%%%%
\section{\label{TeleGraS}Teleparallel Gravity with a Scalar}

We will study the theories of teleparallel gravity coupled with
a scalar in which gravity is  described by a dynamical scalar
$\phi$ in addition to the vierbein $\tensor{e}{_a^\mu}$.
Without loss of generality, we consider the Brans-Dicke-like
theories, whose actions are given by
\be\label{action}
	S = \frac{1}{2 \kappa^2} \int dx^4 e \left[
			\phi T
			- \frac{\omega(\phi)}{\phi} (\partial \phi)^2
			- 2 \kappa^2 V(\phi)
			\right]
		+ S_m\left[ \tensor{e}{_a^\mu}, \chi_m \right] .
\ee
where the coupling function $\omega(\phi)$ and the potential
$V(\phi)$ are two arbitrary functions of $\phi$. At first
glance, one might consider that this action is not so general.
In fact, we can make it more familiar. Introducing a new scalar
$\tilde{\phi}$ according to $( \partial \tilde{\phi} )^2 =
-\omega(\phi) ( \partial\phi )^2 / ( \kappa^2 \phi )$,
Eq.~(\ref{action}) can be recast as
\be\label{actionnewphi}
 \tilde{S}
 = \int dx^4 e \left[
 		\tilde{\omega}(\tilde{\phi}) \frac{T}{2 \kappa^2}
		+ \frac{1}{2} ( \partial \tilde{\phi} )^2
		- \tilde{V}(\tilde{\phi})
		\right]
		+ S_m\left[ \tensor{e}{_a^\mu}, \chi_m \right] .
\ee
Obviously, if $\tilde{\omega}(\tilde{\phi}) =
1 + \xi \kappa^2 \tilde{\phi}^2$, Eq. \eqref{actionnewphi}
reduces to the action considered in \cite{Li:2013oef}, namely
\be\label{actionteleDE}
 \tilde{S}
 = \int dx^4 e \left\{
 	\frac{T}{2 \kappa^2} + \frac{1}{2} \left[
 ( \partial \tilde{\phi} )^2 +\xi\, T \tilde{\phi}^2 \right]
		- \tilde{V}(\tilde{\phi})
		\right\}
		+ S_m\left[ \tensor{e}{_a^\mu}, \chi_m \right] .
\ee
So, the action \eqref{action} is general enough in fact (see
Sec.~\ref{Conclusions} for further discussion). Actually, the
action \eqref{actionnewphi} has been considered
in e.g. \cite{Otalora:2013tba,Kucukakca:2013mya,Izumi:2013dca}
as a generalization of the action \eqref{actionteleDE}. We
stress that the action \eqref{actionnewphi} has richer
structure and more physical implication than the action
\eqref{actionteleDE}, thus justifying the worth of our work.
For instance, it is claimed in \cite{Otalora:2013tba} that the
action \eqref{actionnewphi} might admit the scaling attractors
to alleviate the cosmological coincidence problem, while no
scaling attractor has been found by performing dynamical
analysis of the action \eqref{actionteleDE} (see e.g.
\cite{Wei:2011yr,Xu:2012jf}). On the other hand, using Lagrange
multiplier, $f(T)$ gravity can be recast in a form like the
action \eqref{action} with $\omega(\phi)=0$ (see
e.g. \cite{Yang:2010ji}). Despite the action \eqref{action} is
more general than the action \eqref{actionteleDE} because it
can also encompass $f(T)$ theory when $\omega(\phi)=0$, we will
not consider the case of $\omega(\phi)=0$ in this work, since
then $\phi$ will not be a dynamical quantity.

The variation of the action \eqref{action}
with respect to the scalar field $\phi$ yields
\be\label{eom_phi}
	T
	- \left( \frac{\omega'}{\phi} + \frac{\omega}{\phi^2} \right)
		(\partial \phi)^2
	+ \frac{2}{\phi} \partial_\mu \omega \partial^\mu \phi
	+ 2 \frac{\omega}{\phi} \Box \phi
	- 2 \kappa^2 V'
	= 0,
\ee
where a prime denotes a derivative with respect to $\phi$, and
$\Box = g^{\mu \nu} \nabla_\mu \nabla_\nu$ is the d'Alembert
operator, with $\nabla_\mu$ the covariant derivative associated
with the Levi-Civita connection. The variation of the
action \eqref{action} with respect to the dual vierbein
$\tensor{e}{^a_\nu}$ yields
\be\label{eom00}
	\phi\, e^{-1} \partial_{\sigma} ( e \tensor{S}{_a^\sigma^\nu} )
		+ \tensor{S}{_a^\sigma^\nu} \partial_{\sigma} \phi
		+ \phi \left(
			\tensor{T}{^\sigma_\alpha_a}
			\tensor{S}{_\sigma^\nu^\alpha}
			+ \frac{T}{4} \tensor{e}{_a^\nu}
			\right)
		- \frac{1}{4} \tensor{e}{_a^\nu} \left[
			\frac{\omega}{\phi} (\partial \phi)^2
			+ 2 \kappa^2 V
			\right]
	= \frac{\kappa^2}{2} \tensor{\mathcal{T}}{_a^\nu} ,
\ee
where $\tensor{\mathcal{T}}{_a^\nu} \equiv
- e^{-1}\,\delta S_m / \delta \tensor{e}{^a_\nu}$
is the energy-momentum of matter. Let us bring Eq.~\eqref{eom00} to
a more suitable form for our purpose. Multiplying each side of
Eq.~\eqref{eom00} by the dual vierbein $\tensor{e}{^a_\mu}$, we get
\be\label{eom01}
	\phi\, e^{-1} \tensor{e}{^a_\mu} \partial_{\sigma} \left(
			e \tensor{S}{_a^\sigma^\nu}
			\right)
		+ \tensor{S}{_\mu^\sigma^\nu} \partial_{\sigma} \phi
		+ \phi \left(
			\tensor{T}{^\sigma_\alpha_\mu}
				\tensor{S}{_\sigma^\nu^\alpha}
			+ \frac{T}{4} \delta_\mu^\nu
			\right)
		- \frac{1}{4} \delta_\mu^\nu \left[
			\frac{\omega}{\phi} (\partial \phi)^2
			+ 2 \kappa^2 V
			\right]
	= \frac{\kappa^2}{2} \tensor{\mathcal{T}}{_\mu^\nu} ,
\ee
where we have used the vierbein (or dual vierbein) to switch from
Latin to Greek indices and back, for example $\tensor{\mathcal{T}}
{_\mu^\nu}= \tensor{e}{^a_\mu} \tensor{\mathcal{T}}{_a^\nu}$.
Taking the trace of Eq.~\eqref{eom01} leads to
\be\label{eom02}
	\phi\, e^{-1} \tensor{e}{^a_\rho} \partial_{\sigma} \left(
			e \tensor{S}{_a^\sigma^\rho}
			\right)
		+ \tensor{S}{_\rho^\sigma^\rho} \partial_{\sigma} \phi
		- \left[
			\frac{\omega}{\phi} (\partial \phi)^2
			+ 2 \kappa^2 V
			\right]
	= \frac{\kappa^2}{2} \mathcal{T} ,
\ee
with $\mathcal{T} = \tensor{\mathcal{T}}{_\mu^\mu}$.
Multiplying Eq.~\eqref{eom02} by $\left( -\delta^\nu_\mu / 2
\right)$, then adding Eq.~\eqref{eom01}, we get
\be\label{eom_vierbein}
\begin{split}
	\phi\, e^{-1} \tensor{e}{^a_\mu} \partial_{\sigma} \left(
			e \tensor{S}{_a^\sigma^\nu}
			\right)
		- \frac{1}{2} \delta^{\nu}_{\mu} \phi\, e^{-1}
			\tensor{e}{^a_\rho} \partial_{\sigma} \left(
				e \tensor{S}{_a^\sigma^\rho}
				\right)
		+ \tensor{S}{_\mu^\sigma^\nu} \partial_{\sigma} \phi
		- \frac{1}{2} \delta^{\nu}_{\mu} \tensor{S}{_\rho^\sigma^\rho}
			\partial_{\sigma} \phi \\
		+ \,\phi \left(
			\tensor{T}{^\sigma_\alpha_\mu}
				\tensor{S}{_\sigma^\nu^\alpha}
			+ \frac{T}{4} \delta_\mu^\nu
			\right)
		+ \frac{1}{4} \delta_\mu^\nu \left[
			\frac{\omega}{\phi} (\partial \phi)^2
			+ 2 \kappa^2 V
			\right]
	= \frac{\kappa^2}{2} \left(
			\tensor{\mathcal{T}}{_\mu^\nu}
			- \frac{1}{2} \delta^\nu_\mu \mathcal{T}
			\right) .
\end{split}
\ee
The gravitational fields are truly governed by the field equations
\eqref{eom_phi} and \eqref{eom_vierbein}. We will expand these
two equations in the post-Newtonian approximation in the
following section.

%%%%%%%%%%%%%%%%%%%%%%%%%%%%%%%%%%%%%%%%%%%%%%%%%%%%%%%%%%%%%%%%%%%%%%
%%%%%%%%%%%%%%%%%%%%%%%%%%%%% section 4 %%%%%%%%%%%%%%%%%%%%%%%%%%%%%%
%%%%%%%%%%%%%%%%%%%%%%%%%%%%%%%%%%%%%%%%%%%%%%%%%%%%%%%%%%%%%%%%%%%%%%
\section{\label{PNA}Post-Newtonian Approximation}

The post-Newtonian approximation of GR on the behavior of
hydrodynamic systems has been systematically investigated in
e.g. \cite{chandrasekhar1965post}. In analogy to
\cite{chandrasekhar1965post}, we assume that the gravitating source
matter is contributed by a perfect fluid which obeys the
post-Newtonian hydrodynamics. We will use the PPN formalism to
expand the field equations \eqref{eom_phi} and \eqref{eom_vierbein}
perturbatively by assigning appropriate orders of magnitude to all
dynamical variables appearing in the field equations. The resulting
perturbation equations can then be subsequently solved order by
order.

%%%%%%%%%%%%%%%%%%%%%%%%%%%%%%%%%%%%%%%%%%%%%%%%%%%%%%%%%%%%%%%%%%%%%%
%%%%%%%%%%%%%%%%%%%%%%%%%%%%%%%%%%%%%%%%%%%%%%%%%%%%%%%%%%%%%%%%%%%%%%
\subsection{General framework}
Conventionally, the velocity of the source matter $|\vec{v}|$
characterize the smallness of the system. So, we will
perturbatively expand all dynamical quantities in orders of
$\mathcal{O}(n) \sim |\vec{v}|^n $. We will firstly find out
the perturbations for the vierbein following \cite{Li:2013oef},
and then expand the energy-momentum tensor to sufficient
orders. Finally, the perturbations of all functions of
$\phi$ are obtained by using Taylor expansion.

For the gravitational sector, we expand the dual vierbein
fields around the flat background as
\be
	\tensor{e}{^a_\mu}
	= \tensor{\delta}{^a_\mu} + \tensor{B}{^a_\mu}
	= \tensor{\delta}{^a_\mu}
	  	+ \tensor[^{(1)}]{B}{^a_\mu}
	 	+ \tensor[^{(2)}]{B}{^a_\mu}
	  	+ \tensor[^{(3)}]{B}{^a_\mu}
	  	+ \tensor[^{(4)}]{B}{^a_\mu}
	  	+ \mathcal{O}(5) ,
\ee
where each term $\tensor[^{(n)}]{B}{^a_\mu}$ is of order
$\mathcal{O}(n)$. By using Eq.~\eqref{gTe}, this decomposition
gives the usual metric as an expansion around the flat
Minkowski background,
\be
	g_{\mu \nu}
	= \eta_{\mu \nu} + h_{\mu \nu}
	= \eta_{\mu \nu}
		+ \tensor[^{(1)}]{h}{_\mu_\nu}
		+ \tensor[^{(2)}]{h}{_\mu_\nu}
	  	+ \tensor[^{(3)}]{h}{_\mu_\nu}
	  	+ \tensor[^{(4)}]{h}{_\mu_\nu}
	  	+ \mathcal{O}(5) ,
\ee
where $\eta_{\mu \nu}$ is the Minkowski metric and each symmetric
term $\tensor[^{(n)}]{h}{_\mu_\nu}$ is of order $\mathcal{O}(n)$. For
our purpose, it is sufficient to expand the metric up to the order
of $\mathcal{O}(4)$. A detailed analysis (see e.g.
\cite{will1993theorya}) shows that $\tensor[^{(1)}]{h}{_\mu_\nu} = 0$, which
corresponds to $\tensor[^{(1)}]{B}{^a_\mu} = 0$ (nb. Eq.~(\ref{gTe})),
and the only non-vanishing components of the metric perturbations are
\be
	\tensor[^{(2)}]{h}{_0_0} ,
	\quad \tensor[^{(2)}]{h}{_i_j} ,
	\quad \tensor[^{(3)}]{h}{_0_i} ,
	\quad \tensor[^{(4)}]{h}{_0_0} .
\ee
Following \cite{Li:2013oef}, we denote
$\tensor{B}{^\nu_\mu} = \tensor{\delta}{_a^\nu} \tensor{B}{^a_\mu}$,
or equivalently
$\tensor{B}{^a_\mu} = \tensor{\delta}{^a_\nu} \tensor{B}{^\nu_\mu}$,
with $\tensor{\delta}{^a_\mu}$ defined by $\eta_{\mu \nu} =
\eta_{a b} \tensor{\delta}{^a_\mu} \tensor{\delta}{^b_\nu}$.
We now can raise and lower the spacetime indices of the
perturbations of vierbein (or dual vierbein) by the Minkowski
metric $\eta_{\mu \nu}$,
\be
	B_{\mu \nu} = \eta_{\mu \rho} \tensor{B}{^\rho_\nu} .
\ee
As a result, $B_{\mu \nu}$ is symmetric, and the non-vanishing
components are
\be
	\tensor[^{(2)}]{B}{_0_0} ,
	\quad \tensor[^{(2)}]{B}{_i_j} ,
	\quad \tensor[^{(3)}]{B}{_0_i} ,
	\quad \tensor[^{(4)}]{B}{_0_0} .
\ee
In addition, ${^{(2)}B_{i j}}$ is diagonal \cite{Li:2013oef}.
For convenience, we introduce a time-independent function $A$,
such that ${^{(2)}B_{i j}} = A\delta_{i j}$. We also give the
relations between the metric perturbations and the vierbein
perturbations \cite{Li:2013oef},
\begin{subequations}
\begin{align}
	\tensor[^{(2)}]{h}{_0_0} &= 2\, \tensor[^{(2)}]{B}{_0_0} ,\\
	\tensor[^{(2)}]{h}{_i_j} &= 2\, \tensor[^{(2)}]{B}{_i_j} ,\\
 	\tensor[^{(3)}]{h}{_0_i} &= 2\, \tensor[^{(3)}]{B}{_0_i} ,\\	
 	\tensor[^{(4)}]{h}{_0_0} &= 2\, \tensor[^{(4)}]{B}{_0_0}
 			- \tensor[^{(2)}]{B}{_0_0} \tensor[^{(2)}]{B}{_0_0} .
\end{align}
\end{subequations}
From the definitions, we see that $\tensor{T}{^\rho_\mu_\nu}$ and
$\tensor{S}{^\rho_\mu_\nu}$ are at least $\mathcal{O}(2)$ quantities,
and the torsion scalar $T$ is an at least $\mathcal{O}(4)$ quantity.

The energy-momentum tensor of a perfect fluid takes the form
\be\label{energy-momentum}
	\mathcal{T}^{\mu \nu}
	= ( \rho + \rho \Pi + p ) u^\mu u^\nu
	  + p g^{\mu \nu} ,
\ee
where $\rho$, $\Pi$, $p$ and $u^{\mu}$ are the energy density,
the specific internal energy, the pressure, and the four-velocity of
the fluid, respectively. Note that the velocity of the source matter
is given by $v^i = u^i / u^0$. We assign the velocity orders
$\rho \sim \Pi \sim \mathcal{O}(2)$, and $p \sim \mathcal{O}(4)$ by
considering their orders of magnitude in the Solar System
\cite{will1993theorya}. Then we get the perturbations of
energy-momentum tenor in Eq.~\eqref{energy-momentum} as
\begin{subequations}
\begin{align}
	\tensor{\mathcal{T}}{_0^0}
	&= -\rho \left( 1 + v^2 + \Pi \right) + \mathcal{O}(6) ,\\
	\tensor{\mathcal{T}}{_0^i}
	&= - \rho v^i + \mathcal{O}(5) ,\\
	\tensor{\mathcal{T}}{_i^j}
	&= \rho v_i v^j + p \delta^j_i	+ \mathcal{O}(6) .
\end{align}
\end{subequations}
We also note that $\mathcal{T} = g_{\mu \nu} \mathcal{T}^{\mu \nu}
= -\rho - \rho \Pi + 3 p$. In addition, we assume the gravitational
field  is quasi-static, so the time derivative $\partial_0=\partial/
\partial t$ of the vierbein or other fields are weighted with
an additional velocity order $\mathcal{O}(1)$.

For the scalar field $\phi$, we expand it around its cosmological
background value $\phi_0$,
\be
	\phi = \phi_0 + \psi
	 	 = \phi_0 + \psi^{(2)} + \psi^{(4)} + \mathcal{O}(6) ,
\ee
where we assume $\phi_0$ to be of order $\mathcal{O}(0)$ and
the perturbations $\psi^{(n)}$ are of order $\mathcal{O}(n)$ as
usual. We also need to expand the functions $\omega(\phi)$ and
$V(\phi)$ around $\phi_0$. To this end, we expand them
using Taylor expansion to sufficient orders,
\begin{subequations}
\begin{align}
	\omega &= \omega_0 + \omega_1 \psi
			+ \mathcal{O}\left(\psi^2\right) ,\\
	V &= V_0 + V_1 \psi + V_2 \psi^2 + V_3 \psi^3
			+ \mathcal{O}\left(\psi^4\right) ,
\end{align}
with $\omega_0 = \omega(\phi_0)$, $\omega_1 = \omega'(\phi_0)$,
$V_0 = V(\phi_0)$, $V_1 = V'(\phi_0)$,
$V_2 = \frac{1}{2} V''(\phi_0)$,
and $V_3 = \frac{1}{6} V'''(\phi_0)$.
We assume all these expansion coefficients to be of order
$\mathcal{O}(0)$. We also give the expansion of $\omega'$ and
$V'$ for further convenience,
\begin{align}
	\omega' &= \omega_1 + \mathcal{O}(\psi) ,\\
	V' &= V_1 + 2 V_2 \psi + 3 V_3 \psi^2
			+ \mathcal{O}\left(\psi^3\right) .
\end{align}
\end{subequations}

%%%%%%%%%%%%%%%%%%%%%%%%%%%%%%%%%%%%%%%%%%%%%%%%%%%%%%%%%%%%%%%%%%%%%%
%%%%%%%%%%%%%%%%%%%%%%%%%%%%%%%%%%%%%%%%%%%%%%%%%%%%%%%%%%%%%%%%%%%%%%
\subsection{Solving the perturbed equations}
Here we will solve the perturbed equations order by order.
We refer to Appendix~\ref{PertForT} for a detailed
computation of the corresponding quantities up to
the appropriate orders. In the followings, we just give
the results directly.

Expanding Eqs.~\eqref{eom_vierbein} and \eqref{eom_phi} to
$\mathcal{O}(0)$ simply gives the solutions $V_0 = V_1 = 0$.
We then expand Eq.~\eqref{eom_phi} to $\mathcal{O}(2)$ to get
\be\label{psi02eq}
	\left( \nabla^2 - m_\psi^2 \right) \psi^{(2)} = 0,
\ee
for the scalar field perturbation $\psi^{(2)}$, where
$\nabla^2 = \delta^{i j} \partial_i \partial_j$
and $m_\psi = 2 \kappa \sqrt{\frac{V_2 \phi_0}{2 \omega_0}}$.
Eq.~\eqref{psi02eq} is a screened Poisson equation. Since we
demand that $\phi$ to take its cosmological value at large
scale, which is equivalent to saying that the perturbation
should vanish at cosmological distance due to the absence
of the gravitational field and the matter source, i.e.,
$\psi^{(2)} \rightarrow 0$ as $r \rightarrow \infty$
($r$ is the distance from the Sun), we get the solution of
Eq.~\eqref{psi02eq} as
\be\label{psi02}
	\psi^{(2)} = 0 .
\ee

In order to get the corresponding vierbein perturbations,
we use the ansatz
\be\label{ansatz}
	\tensor[^{(2)}]{h}{_i_j}
		= \gamma(r)\, \tensor[^{(2)}]{h}{_0_0} \delta_{i j}
		= 2 \gamma(r) A \delta_{i j} ,
\ee
where $\gamma(r)$ is a PPN parameter measuring the amount of
space curvature produced by unit rest mass~\cite{will1993theorya}.
We also adopt the gauge conditions for the
vierbein perturbation $\tensor{B}{^\mu_\nu}$ as
\cite{smalley1980post}
\begin{subequations}\label{gauge}
\begin{align}
	\partial_j \tensor[^{(2)}]{B}{_i^j}
		- \frac{1}{2} \partial_i \tensor[^{(2)}]{B}{_\mu^\mu}
	= \frac{1}{2 \phi_0} \partial_i {\psi^{(2)}} = 0 ,\\
	\partial_j \tensor[^{(3)}]{B}{_0^j}
		- \frac{1}{2} \partial_0 \tensor[^{(2)}]{B}{_j^j}
	= \frac{1}{2 \phi_0} \partial_0 {\psi^{(2)}} = 0 ,
\end{align}
\end{subequations}
in which we have used Eq.~\eqref{psi02} actually. These gauge
conditions can directly lead to the standard gauge formulas
\cite{nutku1969post,will1971theoretical}
\begin{subequations}
\begin{align}
	\partial_j \tensor[^{(2)}]{h}{_i^j}
	- \frac{1}{2} \partial_i \tensor[^{(2)}]{h}{_\mu^\mu}
	= \frac{1}{\phi_0} \partial_i {\psi^{(2)}} = 0 ,\\
		\partial_j \tensor[^{(3)}]{h}{_0^j}
		- \frac{1}{2} \partial_0 \tensor[^{(2)}]{h}{_j^j}
	= \frac{1}{\phi_0} \partial_0 {\psi^{(2)}} = 0 .
\end{align}
\end{subequations}
We should verify the consistency of these gauge conditions
after obtaining the solutions. Actually, as we will see later,
our results are identical to GR, so these conditions are just
the Newtonian continuity equations \cite{nutku1969post}, and
are satisfied automatically.

Expanding $(0, 0)$ component of Eq.~\eqref{eom_vierbein} to
$\mathcal{O}(2)$, we get
\be\label{00o2}
	\phi_0 \left[
			\partial_k \tensor{S}{_0^k^0}
			- \frac{1}{2} \partial_k \tensor{S}{_\rho^k^\rho}
			\right]
	= - \frac{1}{4} \kappa^2 \rho
	= \frac{1}{2} \nabla^2 U ,
\ee
in which the gravitational potential $U$ is defined by
\be
	\nabla^2 U = - \frac{1}{2} \kappa^2 \rho .
\ee
The solution to this equation is
\be
	A = \frac{U}{\phi_0} .
\ee

Expanding $(i, j)$ component of Eq.~\eqref{eom_vierbein} to
$\mathcal{O}(2)$, we get
\be\label{ijo21}
	\phi_0 \left[
			\partial_k \tensor{S}{_i^k^j}
			- \frac{1}{2} \delta^j_i \partial_k
				\tensor{S}{_\rho^k^\rho}
			\right]
	= \frac{1}{4} \delta^j_i \kappa^2 \rho .
\ee
Taking the trace of Eq.~\eqref{ijo21} yields
\be\label{ijo22}
	\phi_0 \left[
			\partial_k \tensor{S}{_i^k^i}
			- \frac{3}{2} \partial_k \tensor{S}{_\rho^k^\rho}
			\right]
	= - \frac{3}{2} \nabla^2 U .
\ee
The solution to Eq.~\eqref{ijo22} is given by
\be\label{gammaresult}
	\gamma(r) = 1 .
\ee

Expanding Eq.~\eqref{eom_phi} to $\mathcal{O}(4)$ yields
\be
	T
		+ 2 \frac{\omega_0}{\phi_0} \nabla^2 \psi^{(4)}
		- 4 \kappa^2 V_2 \psi^{(4)}
	= 0 .
\ee
Noting that $T = 2 \partial_i A \partial^i A$ (see
Eq.~\eqref{To4}), the above equation can be simplified to
\be\label{eomphi4}
	\left( \nabla^2 - m_\psi^2 \right) \psi^{(4)}
	= \frac{1}{\phi_0 \omega_0} \nabla^2
		\left( \Phi_2 - \frac{U^2}{2} \right) ,
\ee
where we have used the identity
\be
	\partial_i U \partial^i U
	= \frac{1}{2} \nabla^2 U^2 - U \nabla^2 U ,
\ee
and $\Phi_2$ is defined by
\be
	\nabla^2 \Phi_2 = - \frac{\kappa^2}{2} \rho U .
\ee
Eq.~\eqref{eomphi4} is a screened Poisson equation and
can be solved by
\be\label{phi4}
	\psi^{(4)} = \frac{1}{\phi_0 \omega_0}
		\left( \Phi_2 - \frac{U^2}{2} \right) e^{- m_\psi r} .
\ee

Expanding $(0, i)$ component of Eq.~\eqref{eom_vierbein} to
$\mathcal{O}(3)$, we obtain
\be
	\phi_0 e^{-1} \tensor{e}{^a_0} \partial_\sigma
			\left( e \tensor{S}{_a^\sigma^i} \right)
	= \frac{\kappa^2}{2} \left( - \rho v^i \right) .
\ee
The solution to this equation is
\be\label{B3}
	\tensor[^{(3)}]{B}{_0_i}	
	= - \frac{1}{\phi_0} \left(
			\frac{7}{4} \mathcal{V}_i
			+ \frac{1}{4} W_i
			\right) ,
\ee
with $\mathcal{V}_i$ and $W_i$ defined as in
\cite{chandrasekhar1965post},
\be
	 \nabla^2 \mathcal{V}_i
	 = - \frac{\kappa^2}{2} \rho v_i ,
\ee
and
\be
	W_i
	= G_N \int d^3 y\,
			\frac{\rho(y,t) v^k(y,t) (x-y)_k (x-y)_i}{|x-y|^3} .
\ee
Note that we have used the fact $2 \partial_0 \partial_i (\phi_0 A)
= - \nabla^2 (\mathcal{V}_i - W_i)$ \cite{chandrasekhar1965post} to
derive Eq.~\eqref{B3}.

Expanding $(0, 0)$ component of Eq.~\eqref{eom_vierbein} to
$\mathcal{O}(4)$, we obtain
\be
	\phi_0 e^{-1} \tensor{e}{^a_0} \partial_\sigma
			\left( e \tensor{S}{_a^\sigma^0} \right)
		- \frac{1}{2} \phi_0 e^{-1} \tensor{e}{^a_\rho}
			\partial_\sigma \left( e \tensor{S}{_a^\sigma^\rho} \right)
		+ \phi_0 \left(
			\tensor{T}{^\sigma_\alpha_0} \tensor{S}{_\sigma^0^\alpha}
			+ \frac{T}{4}
			\right)
	= \frac{\kappa^2}{2} \left(
			\tensor{\mathcal{T}}{_0^0}
			- \frac{1}{2} \mathcal{T}
			\right) .
\ee
The solution to this equation is
\be
	B_{0 0}
	= \frac{1}{\phi_0} ( U + 2 \Phi_1 + \Phi_3 + 3 \Phi_4 )
		+ \frac{2}{\phi_0^2} \Phi_2
	  	- \frac{1}{2 \phi_0^2} U^2 ,
\ee
where $\Phi_1$, $\Phi_3$, and $\Phi_4$ are defined as in
\cite{will1993theorya},
\begin{subequations}
\begin{align}
	\nabla^2 \Phi_1 &= - \frac{\kappa^2}{2} \rho v^2 ,\\
	\nabla^2 \Phi_3 &= - \frac{\kappa^2}{2} \rho \Pi ,\\
	\nabla^2 \Phi_4 &= - \frac{\kappa^2}{2} p .
\end{align}
\end{subequations}

In summary, we get the corresponding metric perturbations as
\begin{subequations}
\begin{align}
	{h_{0 0}} &= \frac{2}{\phi_0} ( U + 2 \Phi_1 + \Phi_3 + 3 \Phi_4 )
	 		+ \frac{4}{\phi_0^2} \Phi_2
	 		- \frac{2}{\phi_0^2} U^2 ,\\
	{h_{0 j}} &= - \frac{2}{\phi_0} \left(
				\frac{7}{4} {\cal V}_i
				+ \frac{1}{4} W_i
				\right) ,\\
	{h_{i j}} &= 2 \frac{U}{\phi_0} \delta_{i j} .
\end{align}
\end{subequations}
From above equations, it is easy to see that the effective
Newtonian constant $G_{\rm eff} = G_N /\phi_0$, and the PPN
parameter $\beta(r)$ is given by
\be\label{betaresult}
\beta(r) = 1.
\ee
We note that the PPN parameter $\beta(r)$ measures the amount
of ``non-linearity'' in the superposition law for gravity
\cite{will1993theorya}. Notice that Eqs.~(\ref{gammaresult})
and (\ref{betaresult}) are the main results of this work.

%%%%%%%%%%%%%%%%%%%%%%%%%%%%%%%%%%%%%%%%%%%%%%%%%%%%%%%%%%%%%%%%%%%%%%
%%%%%%%%%%%%%%%%%%%%%%%%%%%%% section 5 %%%%%%%%%%%%%%%%%%%%%%%%%%%%%%
%%%%%%%%%%%%%%%%%%%%%%%%%%%%%%%%%%%%%%%%%%%%%%%%%%%%%%%%%%%%%%%%%%%%%%
\section{\label{Conclusions}Conclusions and discussions}

We have studied the post-Newtonian approximation of teleparallel
gravity coupling to a scalar field $\phi$ with arbitrary
coupling function $\omega(\phi)$ and arbitrary potential $V(\phi)$.
We have chosen frames in which the Sun is at rest in both the
coordinate frame and the tetrad frame, such that the vierbein
(dual vierbein) can be perturbatively expanded around the flat
spacetime, which leads to the usual expanding of the metric
around the Minkowski spacetime. The functions $\omega(\phi)$
and $V(\phi)$ are characterized by the coefficients of Taylor
expansion. Interestingly, the only non-vanishing PPN parameters
$\beta$ and $\gamma$ are all equal to $1$, indicating that
these models are indistinguishable from GR in the Solar System
distance up to the post-Newtonian order. In addition, we can
rescale the cosmological background value $\phi_0$ of the
scalar to $\phi_0 = 1$, and then $G_{\rm eff} = G_N$. Since the
rescaling can be done globally, we conclude that the effective
Newtonian constant has no contribution to the Solar
System experiments neither.

This feature makes the theories we studied quite different
from the scalar-tensor theories (nb.~\cite{Hohmann:2013rba}),
which might be subject to stringent constraints on the
parameter space, or need some screening mechanisms to pass the
Solar System experimental constraints. We might conclude that
the coupling between the scalar field and the torsion scalar
in teleparallel gravity is less strong as that between the
scalar and the Ricci scalar in GR. This can be seen from the
relationship between the torsion scalar constructed from the
Weitzenb\"ock connection and the Ricci scalar constructed from
the Levi-Civita connection \cite{Maluf:1994ji},
\be\label{TandR}
	T = - R - 2 \nabla^\mu \tensor{T}{^\nu_\mu_\nu} .
\ee
Although the second term on the right hand side of
Eq.~\eqref{TandR} is a boundary term in the TEGR case, it will
be nontrivial when a scalar field $\phi$ is coupled to the torsion
\be
	\phi T = - \phi R
			- 2 \phi \nabla^\mu \tensor{T}{^\nu_\mu_\nu} ,
\ee
which makes the theories quite different from the scalar-tensor
theories. In addition, $T$ is at least $\mathcal{O}(4)$, while
$R$ is at least $\mathcal{O}(2)$ when perturbated around the
flat spacetime. This fact makes the gravitational sector have
no effect on the $\psi^{(2)}$ when Eq.~\eqref{eom_phi} is
expanded up to $\mathcal{O}(2)$, thus leading to the PPN
parameter $\gamma(r)$ equals to $1$. This result is agree with
the previous work in \cite{Li:2013oef}. The authors in
\cite{Li:2013oef} have argued that, since the source matter is
not involved in the solution of $\mathcal{O}(2)$ perturbation
of the scalar field (see Eq.~\eqref{eom02} in our case), the
Newtonian potential cannot be modified to a Yukawa type
$U(r) = U e^{-mr}$ as in scalar-tensor theories. Although the
non-minimally coupling between the scalar and the torsion shows
no deviation from GR in the post-Newtonian approximation, the
distinction may appear in the post-post-Newtonian
\cite{Epstein:1980dw} limit when such experiments are available. In
fact, the scalar perturbation $\psi^{(4)}$ is non-vanishing
(see Eq.~\eqref{phi4}), and it definitely will affect the
post-post-Newtonian behavior through Eq.~\eqref{eom_vierbein}.
This indirect coupling between the scalar field and the gravitational
sector is the meaning of less strong coupling we proposed.

Similar to $f(T)$ theory \cite{Li:2010cg}, the action \eqref{action}
is not invariant under local Lorentz transformation.
One might, therefore, expect some preferred-frame
effects to show up in post-Newtonian limit (we thank the
referee for pointing out this issue). Although our results reveal no
coordinate frame is preferred in obtaining PPN parameters, there is
indeed a preferred tetrad frame in our calculation. It is
interesting to note that similar results have been achieved in some
scalar-tetrad theories of gravity (see e.g.
\cite{Hayward:1979zg,Hayward:1981bk} and the references therein).
It is claimed in \cite{Hayward:1979zg,Hayward:1981bk} that the
preferred-tetrad-frame effect cannot be detected
if one only measures the metric components.
Attempts to measure the tetrad in a direct way, e.g., the
interaction of tetrad with a spin-1/2 field, would
generally introduce some Lorentz gauge fields
to restore the local Lorentz symmetry \cite{hehl1980gravitation},
and thus creating a Poincar\'{e} gauge
theory. We refer to \cite{Hayward:1979zg,Hayward:1981bk} for a more
detailed discussion of this issue.

We stress here that not all kinds of non-minimally coupling
between the torsion and the scalar would have no affect on the
weak field behavior of the theory (we are indebted to an
anonymous colleague for pointing out this issue).
For example, if we add a term of the form
$\tensor{T}{^\alpha_\alpha_\beta} \partial^\beta \phi$ as
considered in e.g. \cite{saez1983variational,saez1985parametrized}
to the action~\eqref{action}, an extra term like
\be
	- \partial^k \tensor{T}{^\alpha_\alpha_\beta}
\ee
would be added to the $\mathcal{O}(2)$ perturbative equation
of $\phi$ (i.e. Eq.~\eqref{psi02eq}).
Therefore, the value of $\psi^{(2)}$
will not vanish in this case, thus changing the gauge
conditions \eqref{gauge}. So, quite contrary to our original
action \eqref{action}, the additional non-minimally coupling
term $\tensor{T}{^\alpha_\alpha_\beta} \partial^\beta \phi$
might make the PPN parameters differ from the case of GR.
We leave this issue to the future works.

One might note that the action~\eqref{action} considered in this
work could be further generalized to
\be\label{actiongenxi}
	S = \frac{1}{2 \kappa^2} \int dx^4 e \left[
			\xi(\phi)\, T
			- \frac{\omega(\phi)}{\phi} (\partial \phi)^2
			- 2 \kappa^2 V(\phi)
			\right]
		+ S_m\left[ \tensor{e}{_a^\mu}, \chi_m \right] .
\ee
However, it is an illusion. Introducing a new scalar
$\hat{\phi} = \xi(\phi)$, Eq.~(\ref{actiongenxi}) can be recast as
\be\label{actiongenxinewphi}
	\hat{S}
	= \frac{1}{2 \kappa^2} \int dx^4 e \left[
			\hat{\phi}\,T
 			-\frac{\hat{\omega}(\hat{\phi})}{\hat{\phi}}
 				( \partial \hat{\phi} )^2
			- 2 \kappa^2 \hat{V}(\hat{\phi})
			\right]
		+ S_m\left[ \tensor{e}{_a^\mu}, \chi_m \right] ,
\ee
which reduces to the action~\eqref{action} actually. So, the
conclusions do not change for the action~\eqref{actiongenxi}.
This indicates that the action~\eqref{action} considered in
this work is general enough.

Although the theories we studied here have the same PPN
parameters as GR, it differs from GR in several aspects.
Firstly, the deviation from GR might show up in the higher
order perturbation, e.g. in the post-post-Newtonian limit
\cite{Epstein:1980dw}. Secondly, we should consider the
preferred tetrad frame effect  (we thank the referee for
pointing out this issue). Unfortunately, there are no PPN
parameters to characterize this effect. So, we get the same
PPN parameters as GR. The standard post-Newtonian formalism
might be generalized to incorporate this effect. And it is
beyond the scope of the present work.

Finally, from the viewpoint of symmetry, black holes have
similar environments like the Solar System. So, we might
speculate that our theories will have the same solutions as GR when
applying to black holes. Thus, it would be interesting to study
the black hole solutions in the future works.

%%%%%%%%%%%%%%%%%%%%%%%%%%%%%%%%%%%%%%%%%%%%%%%%%%%%%%%%%%%%%%%%%%%%%%
%%%%%%%%%%%%%%%%%%%%%%%%%% acknowledgements %%%%%%%%%%%%%%%%%%%%%%%%%%
%%%%%%%%%%%%%%%%%%%%%%%%%%%%%%%%%%%%%%%%%%%%%%%%%%%%%%%%%%%%%%%%%%%%%%
\begin{acknowledgments}
We thank the anonymous referee for quite useful comments and
suggestions, which helped us to improve this work.
We are indebted to an anonymous colleague for pointing out the
references \cite{saez1983variational,saez1985parametrized}.
We are grateful to Jing Liu, Xiao-Peng~Yan, Ya-Nan~Zhou,
Xiao-Bo~Zou, and Hong-Yu~Li for kind help and discussions.
This work was supported in part by NSFC under
Grants No.~11175016 and No.~10905005, as well as NCET under
Grant No.~NCET-11-0790.
\end{acknowledgments}

\vspace{4mm} % used here just for a better typesetting

%%%%%%%%%%%%%%%%%%%%%%%%%%%%%%%%%%%%%%%%%%%%%%%%%%%%%%%%%%%%%%%%%%%%%%
%%%%%%%%%%%%%%%%%%%%%%%%%%%% appendix %%%%%%%%%%%%%%%%%%%%%%%%%%%%%%%%
%%%%%%%%%%%%%%%%%%%%%%%%%%%%%%%%%%%%%%%%%%%%%%%%%%%%%%%%%%%%%%%%%%%%%%
\appendix 	%% add * for only one appedndix %%
%%%%%%%%%%%%%%%%%%%%%%%%%%%%%%%%%%%%%%%%%%%%%%%%%%%%%%%%%%%%%%%%%%%%%%
\section{\label{PertForT}Perturbations for the torsion tensor
and the~super-potential tensor}

We present here the detailed calculations of the perturbations
for the torsion tensor $\tensor{T}{^\lambda_\mu_\nu}$ and the
super-potential tensor $\tensor{S}{^\rho_\mu_\nu}$ which are
constructed from the vierbein $\tensor{e}{_a^\mu}$ and dual
vierbein $\tensor{e}{^a_\mu}$ to sufficient order.
Note that the ansatz \eqref{ansatz} is equivalent to
\be
	\tensor[^{(2)}]{B}{_i_j} = \gamma A \delta_{i j} , ~~~~~~~
	\tensor[^{(2)}]{B}{_0_0} = A .
\ee
After solving the $(i, j)$ component of Eq.~\eqref{eom_phi} to
$\mathcal{O}(2)$, we have the PPN parameter $\gamma(r)$ equal
to $1$ (nb. Eq.~(\ref{gammaresult})). So, when dealing with
$\mathcal{O}(2)$ quantities, we explicitly show the parameter
$\gamma$ in the expansion, and simplify $\mathcal{O}(3)$
and $\mathcal{O}(4)$ quantities by setting $\gamma = 1$.

We firstly expand the vierbein fields around the flat background as
\be
	\tensor{e}{_a^\mu}
	= \tensor{\delta}{_a^\mu}
		+ \tensor{C}{_a^\mu}
	= \tensor{\delta}{_a^\mu}
	  	+ \tensor[^{(2)}]{C}{_a^\mu}
	  	+ \tensor[^{(3)}]{C}{_a^\mu}
	  	+ \tensor[^{(4)}]{C}{_a^\mu}
	  	+ \mathcal{O}(5),
\ee
where each term $\tensor[^{(n)}]{C}{_a^\mu}$ is of order
$\mathcal{O}(n)$. Noting that
$g^{\mu \nu}(x) = \eta^{a b} \tensor{e}{_a^\mu}(x) \tensor{e}{_b^\nu}(x)$
and using Eq.~\eqref{gTe}, we can easily get
$\tensor[^{(2)}]{C}{_a^\lambda} = - \tensor[^{(2)}]{B}{^\lambda_a}$.
We then expand the torsion tensor $\tensor{T}{^\lambda_\mu_\nu}$
up to $\mathcal{O}(4)$,
\be\label{T_pert}
\begin{split}
	\tensor{T}{^\lambda_\mu_\nu}
	&= \tensor{e}{_a^\lambda} (
			\partial_\nu \tensor{e}{^a_\mu}
			- \partial_\mu \tensor{e}{^a_\nu}
			) \\
	&= (
			\tensor{\delta}{_a^\lambda}
			+ \tensor{C}{_a^\lambda}
			) (
				\partial_\nu \tensor{B}{^a_\mu}
				- \partial_\mu \tensor{B}{^a_\nu}
				) \\
	&= \partial_\nu \tensor{B}{^\lambda_\mu}
		- \partial_\mu \tensor{B}{^\lambda_\nu}
		+ \tensor{C}{_a^\lambda} \partial_\nu \tensor{B}{^a_\mu}
		- \tensor{C}{_a^\lambda} \partial_\mu \tensor{B}{^a_\nu} \\
	&= \partial_\nu \tensor{B}{^\lambda_\mu}
		- \partial_\mu \tensor{B}{^\lambda_\nu}
		- \tensor[^{(2)}]{B}{^\lambda_\alpha}
			\partial_\nu \tensor[^{(2)}]{B}{^\alpha_\mu}
		+ \tensor[^{(2)}]{B}{^\lambda_\alpha}
			\partial_\mu \tensor[^{(2)}]{B}{^\alpha_\nu} .
\end{split}
\ee
For convenience, we also present the definition of the
super-potential tensor $\tensor{S}{^\rho_\mu_\nu}$ here,
\be
	\tensor{S}{^\rho_\mu_\nu}
	= \frac{1}{4} (
			\tensor{T}{^\rho_\mu_\nu}
			- \tensor{T}{_\mu_\nu^\rho}
			+ \tensor{T}{_\nu_\mu^\rho} )
		+ \frac{1}{2} \delta^\rho_\mu \tensor{T}{^\sigma_\nu_\sigma}
		- \frac{1}{2} \delta^\rho_\nu \tensor{T}{^\sigma_\mu_\sigma} .
\ee
In addition, we use the anti-symmetric properties of the torsion
tensor $\tensor{T}{^\lambda_\mu_\nu}$ and the super-potential tensor
$\tensor{S}{^\rho_\mu_\nu}$ to simplify our calculations. Since the
space-space component of metric $g_{i j}$ is expanded around
the usual Euclidean metric $\delta_{i j}$, we do not distinguish the
upper indices and the lower indices of the perturbation quantities
up to appropriate order. Instead, we use the upper indices and the
lower indices interchangeably, e.g. $\tensor[^{(2)}]{B}{^i_j}
= \tensor[^{(2)}]{B}{_i_j} = \tensor[^{(2)}]{B}{_i^j}
= \tensor[^{(2)}]{B}{^i^j}$, up to $\mathcal{O}(2)$.

%%%%%%%%%%%%%%%%%%%%%%%%%%%%%%%%%%%%%%%%%%%%%%%%%%%%%%%%%%%%%%%%%%%%%%
%%%%%%%%%%%%%%%%%%%%%%%%%%%%%%%%%%%%%%%%%%%%%%%%%%%%%%%%%%%%%%%%%%%%%%
\subsection{Up to \texorpdfstring{$\mathcal{O}(2)$}{TEXT}}

The expansion of torsion tensor to $\mathcal{O}(2)$ can be read from
Eq.~\eqref{T_pert} as
\be
	\tensor{T}{^\lambda_\mu_\nu}
	= \partial_\nu \tensor[^{(2)}]{B}{^\lambda_\mu}
		- \partial_\mu \tensor[^{(2)}]{B}{^\lambda_\nu} .
\ee
Some of its components can be obtained directly,
\bea
	 \tensor{T}{^0_i_0} &= \partial_0 \tensor[^{(2)}]{B}{^0_i}
	 		- \partial_i \tensor[^{(2)}]{B}{^0_0}
	 	&= \partial_i A ,\\
	 \tensor{T}{^k_0_0} &= \partial_0 \tensor[^{(2)}]{B}{^k_0}
	 		- \partial_0 \tensor[^{(2)}]{B}{^k_0}
	 	&= 0 ,\\
	 \tensor{T}{^0_i_j} &= \partial_j \tensor[^{(2)}]{B}{^0_i}
	 		- \partial_i \tensor[^{(2)}]{B}{^0_j}
	 	&= 0 ,\\
	 \tensor{T}{^k_j_i} &= \partial_i \tensor[^{(2)}]{B}{^k_j}
	 		- \partial_j \tensor[^{(2)}]{B}{^k_i}
	 	&= \delta^k_j \partial_i \left(\gamma A\right)
	 		- \delta^k_i \partial_j \left(\gamma A\right) ,\\
	 \tensor{T}{^j_i_j} &= \partial_j \tensor[^{(2)}]{B}{^j_i}
	 		- \partial_i \tensor[^{(2)}]{B}{^j_j}
	 	&= -2 \partial_i \left(\gamma A\right),\\
	 \tensor{T}{^i_0_j} &= \partial_j \tensor[^{(2)}]{B}{^i_0}
	 		- \partial_0 \tensor[^{(2)}]{B}{^i_j}
	 	&= 0 .
\eea
And the expansion for some components of the super-potential
tensor $\tensor{S}{^\rho_\mu_\nu}$ is also obtained,
\be
\begin{split}
	\tensor{S}{^0_i_0}
	&= \frac{1}{4} (
			\tensor{T}{^0_i_0}
			- \tensor{T}{_i_0^0}
			+ \tensor{T}{_0_i^0}
			)
		- \frac{1}{2} \tensor{T}{^\sigma_i_\sigma} \\
	&= \frac{1}{4} (
			\tensor{T}{^0_i_0}
			+ \tensor{T}{^i_0_0}
			+ \tensor{T}{^0_i_0}
			)
		- \frac{1}{2} ( \tensor{T}{^0_i_0} + \tensor{T}{^j_i_j} )
	= - \frac{1}{2} \tensor{T}{^j_i_j}
	= \partial_i \left(\gamma A\right) ,
\end{split}
\ee
\vspace{-4mm} % used here just for a better typesetting
\be
\begin{split}
\tensor{S}{^j_i_j}
	&= \frac{1}{4} (
			\tensor{T}{^j_i_j}
			- \tensor{T}{_i_j^j}
			+ \tensor{T}{_j_i^j}
			)
		+ \frac{1}{2} \delta^j_i \tensor{T}{^\sigma_j_\sigma}
		- \frac{1}{2} \delta^j_j \tensor{T}{^\sigma_i_\sigma}
	= \frac{1}{4} (
			\tensor{T}{^j_i_j}
			- \tensor{T}{^i_j_j}
			+ \tensor{T}{^j_i_j}	
			)
		+ \frac{1}{2} \tensor{T}{^\sigma_i_\sigma}
		- \frac{3}{2} \tensor{T}{^\sigma_i_\sigma} \\
	&= \frac{1}{2} \tensor{T}{^j_i_j}
		- ( \tensor{T}{^0_i_0} + \tensor{T}{^j_i_j} )
	= - \frac{1}{2} \tensor{T}{^j_i_j}
		- \tensor{T}{^0_i_0}
	= \partial_i \left(\gamma A \right) - \partial_i A ,
\end{split}
\ee
\vspace{-3mm} % used here just for a better typesetting
\be
\begin{split}
	\tensor{S}{^i_0_j}
	&= \frac{1}{4} (
			\tensor{T}{^i_0_j}
			- \tensor{T}{_0_j^i}
			+ \tensor{T}{_j_0^i}
			)
			- \frac{1}{2} \delta^i_j \tensor{T}{^\sigma_0_\sigma}
	= \frac{1}{4} (
			\tensor{T}{^i_0_j}
			+ \tensor{T}{^0_j_i}
			+ \tensor{T}{^j_0_i}
			)
		- \frac{1}{2} \delta^i_j (
			\tensor{T}{^0_0_0}
			+ \tensor{T}{^k_0_k}
			)
	= 0 .
\end{split}
\ee
We also present the result of $\partial_\mu e$ up to
$\mathcal{O}(2)$ here,
\be
	\partial_\mu e
	= \partial_\mu \sqrt{-g}
	= \partial_\mu \left(
			1
			+ \frac{1}{2}\, \tensor[^{(2)}]{h}{^\nu_\nu}
			\right)
	= \frac{1}{2} \partial_\mu \tensor[^{(2)}]{h}{^\nu_\nu}
	= \partial_\mu ( 3 \gamma A - A ) ,
\ee
where we have used the fact that
\be
	\tensor[^{(2)}]{h}{^\nu_\nu}
	= \tensor[^{(2)}]{h}{^0_0}
		+ \tensor[^{(2)}]{h}{^i_i}
	= - \tensor[^{(2)}]{h}{_0_0}
		+ \tensor[^{(2)}]{h}{_i_i}
	= - 2\, \tensor[^{(2)}]{B}{_0_0}
		+ 2\, \tensor[^{(2)}]{B}{_i_i}
	= - 2 A + 6 \gamma A .
\ee

%%%%%%%%%%%%%%%%%%%%%%%%%%%%%%%%%%%%%%%%%%%%%%%%%%%%%%%%%%%%%%%%%%%%%%
%%%%%%%%%%%%%%%%%%%%%%%%%%%%%%%%%%%%%%%%%%%%%%%%%%%%%%%%%%%%%%%%%%%%%%
\subsection{Up to \texorpdfstring{$\mathcal{O}(3)$}{TEXT}}
The expansion of torsion tensor to $\mathcal{O}(3)$ can be read from
Eq.~\eqref{T_pert} as
\be
	\tensor{T}{^\lambda_\mu_\nu}
	= \partial_\nu {\tensor{B}{^\lambda_\mu}}
		- \partial_\mu {\tensor{B}{^\lambda_\nu}} .
\ee
Some of its components read
\be
\begin{split}
	\tensor{T}{^0_i_j}
	&= \partial_j \tensor{B}{^0_i}
		- \partial_i \tensor{B}{^0_j}
	= \partial_j \tensor[^{(3)}]{B}{^0_i}
		- \partial_i \tensor[^{(3)}]{B}{^0_j} ,
\end{split}
\ee
\vspace{-7mm} % used here just for a better typesetting
\be
\begin{split}
	\tensor{T}{^i_j_0}
	&= \partial_0 \tensor{B}{^i_j}
		- \partial_j \tensor{B}{^i_0}
	= \partial_0 \tensor[^{(2)}]{B}{^i_j}
		- \partial_j \tensor[^{(3)}]{B}{^i_0}
	= \delta^i_j \partial_0 A
		- \partial_j \tensor[^{(3)}]{B}{^i_0} ,
\end{split}
\ee
\vspace{-7mm} % used here just for a better typesetting
\be
\begin{split}
	\tensor{T}{^i_i_0}
	&= 3 \partial_0 A
		- \partial_i \tensor[^{(3)}]{B}{^i_0}
	= \frac{3}{2} \partial_0 A .
\end{split}
\ee
When we derive above equations, the gauge conditions \eqref{gauge}
have been used. Some components of the super-potential read
\be
\begin{split}
	\tensor{S}{^0_i_j}
	&= \frac{1}{4} (
			\tensor{T}{^0_i_j}
			- \tensor{T}{_i_j^0}
			+ \tensor{T}{_j_i^0}
			)
	= \frac{1}{4} (
			\tensor{T}{^0_i_j}
			+ \tensor{T}{^i_j_0}
			- \tensor{T}{^j_i_0}
			)
	= \frac{1}{2} (
			\partial_j \tensor[^{(3)}]{B}{^0_i}
 			- \partial_i \tensor[^{(3)}]{B}{^0_j}
 			) ,
\end{split}
\ee
\vspace{-6mm} % used here just for a better typesetting
\be
\begin{split}
	\tensor{S}{^i_0_i}
	&= \frac{1}{4} (
			\tensor{T}{^i_0_i}
			- \tensor{T}{_0_i^i}
			+ \tensor{T}{_i_0^i}
			)
		- \frac{1}{2} \delta^i_i \tensor{T}{^\sigma_0_\sigma} \\
	&= \frac{1}{4} (
			\tensor{T}{^i_0_i}
			+ \tensor{T}{^0_i_i}
			+ \tensor{T}{^i_0_i}
			)
		- \frac{3}{2} ( \tensor{T}{^0_0_0} + \tensor{T}{^i_0_i} )
	= \tensor{T}{^i_i_0}
	= \frac{3}{2} \partial_0 A .
\end{split}
\ee

%%%%%%%%%%%%%%%%%%%%%%%%%%%%%%%%%%%%%%%%%%%%%%%%%%%%%%%%%%%%%%%%%%%%%%
%%%%%%%%%%%%%%%%%%%%%%%%%%%%%%%%%%%%%%%%%%%%%%%%%%%%%%%%%%%%%%%%%%%%%%
\subsection{Up to \texorpdfstring{$\mathcal{O}(4)$}{TEXT}}

The expansion of torsion tensor to $\mathcal{O}(4)$ can be read from
Eq.~\eqref{T_pert} as
\be
	\tensor{T}{^\lambda_\mu_\nu}
	= \partial_\nu \tensor{B}{^\lambda_\mu}
		- \partial_\mu \tensor{B}{^\lambda_\nu}
	  	- \tensor[^{(2)}]{B}{^\lambda_\alpha}
 			\partial_\nu \tensor[^{(2)}]{B}{^\alpha_\mu}
 	  	+ \tensor[^{(2)}]{B}{^\lambda_\alpha}
 			\partial_\mu \tensor[^{(2)}]{B}{^\alpha_\nu} ,
\ee
which can directly lead to
\be
\begin{split}
	\tensor{T}{^0_i_0}
 	&= \partial_0 \tensor{B}{^0_i}
 		- \partial_i \tensor{B}{^0_0}
 		- \tensor[^{(2)}]{B}{^0_\alpha}
 			\partial_0 \tensor[^{(2)}]{B}{^\alpha_i}
 		+ \tensor[^{(2)}]{B}{^0_\alpha}
 	    	\partial_i \tensor[^{(2)}]{B}{^\alpha_0} \\
 	&= \partial_0 \tensor[^{(3)}]{B}{^0_i}
 		- \partial_i \tensor{B}{^0_0}
 		+ \tensor[^{(2)}]{B}{^0_0} \partial_i
 			\tensor[^{(2)}]{B}{^0_0}
 	= \partial_0 \tensor[^{(3)}]{B}{^0_i}
		- \partial_i \tensor{B}{^0_0}
		+ A \partial_i A ,
\end{split}
\ee
\vspace{-4.1mm} % used here just for a better typesetting
\be
\begin{split}
	\tensor{T}{^j_i_j}
	&= \partial_j \tensor{B}{^j_i}
		- \partial_i \tensor{B}{^j_j}
		- \tensor[^{(2)}]{B}{^j_\alpha}
			\partial_j \tensor[^{(2)}]{B}{^\alpha_i}
		+ \tensor[^{(2)}]{B}{^j_\alpha}
			\partial_i \tensor[^{(2)}]{B}{^\alpha_j} \\
	&= \partial_j \tensor[^{(2)}]{B}{^j_i}
		- \partial_i \tensor[^{(2)}]{B}{^j_j}
		- \tensor[^{(2)}]{B}{^j_k} \partial_j
			\tensor[^{(2)}]{B}{^k_i}
		+ \tensor[^{(2)}]{B}{^j_k} \partial_i
			\tensor[^{(2)}]{B}{^k_j}
	= - 2 \partial_i A
		+ 2 A \partial_i A .
\end{split}
\ee
The components of the super-potential for our interest are also given,
\be
\begin{split}
	\tensor{S}{^0_i_0}
	&= \frac{1}{4} (
			\tensor{T}{^0_i_0}
			- \tensor{T}{_i_0^0}
			+ \tensor{T}{_0_i^0}
			)
		- \frac{1}{2} \tensor{T}{^\sigma_i_\sigma} \\
	&= \frac{1}{4} ( \tensor{T}{^0_i_0} + \tensor{T}{^0_i_0} )
		- \frac{1}{2} ( \tensor{T}{^0_i_0} + \tensor{T}{^j_i_j} )
	= - \frac{1}{2} \tensor{T}{^j_i_j}
	= \partial_i A - A \partial_i A ,
\end{split}
\ee
\vspace{-5mm} % used here just for a better typesetting
\be
\begin{split}
\tensor{S}{^i_j_i}
	&= \frac{1}{4} ( \tensor{T}{^i_j_i} - \tensor{T}{_j_i^i}
		+ \tensor{T}{_i_j^i} )
		+ \frac{1}{2} \delta^i_j \tensor{T}{^\sigma_i_\sigma}
		- \frac{1}{2} \delta^i_i \tensor{T}{^\sigma_j_\sigma}
	= \frac{1}{4} ( \tensor{T}{^i_j_i} + \tensor{T}{^i_j_i} )
		+ \frac{1}{2} \tensor{T}{^\sigma_j_\sigma}
		- \frac{3}{2} \tensor{T}{^\sigma_j_\sigma} \\
	&= \frac{1}{2} \tensor{T}{^i_j_i}
		- ( \tensor{T}{^0_j_0} + \tensor{T}{^i_j_i} )
	= - \frac{1}{2} \tensor{T}{^i_j_i} - \tensor{T}{^0_j_0}
	= - \partial_0 \tensor[^{(3)}]{B}{^0_j}
		+ \partial_j {\tensor{B}{^0_0}}
		- 2 A \partial_j A
		+ \partial_j A .
\end{split}
\ee
Finally, we expand the torsion scalar $T$ up to $\mathcal{O}(4)$ as
\be\label{To4}
\begin{split}
T
	&= \tensor{S}{^\rho_\mu_\nu} \tensor{T}{_\rho^\mu^\nu}
	= \tensor{S}{^\rho_\mu_0} \tensor{T}{_\rho^\mu^0}
		+ \tensor{S}{^\rho_\mu_i} \tensor{T}{_\rho^\mu^i}
	= \tensor{S}{^\rho_0_0} \tensor{T}{_\rho^0^0}
		+ \tensor{S}{^\rho_i_0} \tensor{T}{_\rho^i^0}
		+ \tensor{S}{^\rho_0_i} \tensor{T}{_\rho^0^i}
		+ \tensor{S}{^\rho_j_i} \tensor{T}{_\rho^j^i} \\
	&= \tensor{S}{^0_i_0} \tensor{T}{_0^i^0}
		+ \tensor{S}{^j_i_0} \tensor{T}{_j^i^0}
		+ \tensor{S}{^0_0_i} \tensor{T}{_0^0^i}
		+ \tensor{S}{^j_0_i} \tensor{T}{_j^0^i}
		+ \tensor{S}{^0_j_i} \tensor{T}{_0^j^i}
		+ \tensor{S}{^k_j_i} \tensor{T}{_k^j^i} \\
	&= \tensor{S}{^0_i_0} \tensor{T}{^0_i_0}
		+ \tensor{S}{^0_0_i} \tensor{T}{^0_0_i}
		+ \tensor{S}{^k_j_i} \tensor{T}{^k_j_i}
	= 2 \tensor{S}{^0_i_0} \tensor{T}{^0_i_0}
		+ \tensor{S}{^k_j_i} (
			\delta^k_j \partial_i A
			- \delta^k_i \partial_j A
			) \\
	&= 2 \tensor{S}{^0_i_0} \tensor{T}{^0_i_0}
	= 2 \partial_i A \partial^i A .
\end{split}
\ee

\vspace{7mm} % used here just for a better typesetting

\renewcommand{\baselinestretch}{1.1}

%%%%%%%%%%%%%%%%%%%%%%%%%%%%%%%%%%%%%%%%%%%%%%%%%%%%%%%%%%%%%%%%%%%%%%
%%%%%%%%%%%%%%%%%%%%%%%%%%%%% references %%%%%%%%%%%%%%%%%%%%%%%%%%%%%
%%%%%%%%%%%%%%%%%%%%%%%%%%%%%%%%%%%%%%%%%%%%%%%%%%%%%%%%%%%%%%%%%%%%%%
\bibliography{./ppnts}

\end{document}